\documentclass[12pt]{article}
\usepackage{amsmath,amsfonts,amssymb}
\usepackage{theorem}
\usepackage{graphicx}
\usepackage{color}
\usepackage{longtable}
\usepackage{natbib}
\newcommand{\ra}{\rightarrow}

\newcommand{\Lra}{\Longrightarrow}
\newcommand{\half}{\frac{1}{2}}
\newcommand{\Half}{\frac{3}{2}}

\newcommand{\be}{\begin{equation}}
\newcommand{\ee}{\end{equation}}
\newcommand{\bea}{\begin{eqnarray}}
\newcommand{\eea}{\end{eqnarray}}

\begin{document}


\pagestyle{empty}
 

\begin{flushright}
LAPTH-Conf.036/14
\end{flushright}

\vspace{12mm}

\begin{center}

{\Large \bf{Crystal Basis Model: }

\vspace{2mm}

\bf{Codon-Anticodon  Interaction}

\bf{and}

\vspace{2mm}

\bf{ Genetic Code Evolution}}

\end{center}

\vspace{7mm}
\begin{center}
 {\large A. Sciarrino}

\vspace{4mm}

{\em Dipartimento di Scienze Fisiche, Universit\`a di Napoli
       ``Federico II''}
\\
{\em I.N.F.N., Sezione di Napoli, Italy}
\\
{\em  Complesso Universitario di Monte S. Angelo,

Via Cintia, 80126 Napoli, Italy}
\\
{\em E-mail:} \texttt{sciarrino@na.infn.it}
\vspace{7mm}

{\large P. Sorba}

\vspace{4mm}

{\em Laboratoire de Physique Th\'eorique LAPTH, URA 1436,}
\\
{\em Chemin de Bellevue, BP 110,\\ F-74941 Annecy-le-Vieux, France}
\\
{\em E-mail:} \texttt{sorba@lapp.in2p3.fr}

\vspace{5mm}
\end{center}


\begin{center}
{\textsl Talk presented by P.Sorba at the  \\ TABIS  (Theoretical Approaches 
to BioInformation Systems) 2013 Conference\\ Belgrade, Serbia, Sept.2013}
\end{center}
\vspace{5mm}

%
%
%
%
%
%
%
%
%
%

\begin{abstract}
    Imposing a minimum principle in the framework of the so called crystal basis model of the genetic code, we determine the structure of the minimum set  of 22 anticodons which allows the translational-transcription for animal mitochondrial code. The results are in very good agreement with the observed anticodons. 
 Then, we analyze the evolution of the genetic code, with 20 amino acids encoded from the beginning, from the  viewpoint  of codon-anticodon interaction. Following the same spirit as above, we determine the structure  of the anticodons in the Ancient, Archetypal and Early  Genetic codes.  Most of our results  agree  with the generally accepted scheme. 
\end{abstract}

{\em Keyword}: 
genetic code,  codon, anticodon  


\newpage
\pagestyle{plain}
\setcounter{page}{1}
\baselineskip=16pt

\section{Introduction and Crystal Basis Model}

A few years ago, we proposed a mathematical model, called the ``crystal basis model",  in which the codons appear as composite states of  the four nucleotides. Let us very quickly recall the main ideas of the model introduced in \citep{FSS98}, for a review and some applications see \citep{FSS01}.
   The  
nucleotides, in the following denoted by they first letter (C, U,G, A), being assigned to the fundamental irreducible representation (irrep.) of 
the quantum group ${\cal U}_{q}(su(2) \oplus su(2))$ in the limit $q \to 
0$, the codons are obtained as tensor product of nucleotides.  Indeed, the 
properties of quantum group representations in the limit $q \to 0$, or 
crystal basis, are crucial to take into account the fact that a codon is an ordered triple of nucleotides.
The nucleotide content of the $({\bf \half,\half)}$ (fundamental) representation of ${\cal 
U}_{q \to 0}(su(2) \oplus su(2))$, i.e. the eigenvalues of  $ J_{H,3}, \, J_{V,3}$, is chosen as follows:
\be 
	\mbox{C} \equiv (+\half,+\half) \qquad \mbox{U} \equiv (-\half,+\half) 
	\qquad \mbox{G} \equiv (+\half,-\half) \qquad \mbox{A} \equiv 
	(-\half,-\half)
	\label{eq:gc1}
\ee 
where the first $su(2)$ - denoted  $su(2)_{H}$- corresponds to the distinction between the purine bases A,G and the pyrimidine ones C,U and the second one - denoted $su(2)_{V}$ - corresponds to the complementarity rule C/G and U/A, 
Thus to represent a codon, we have to perform the tensor product of three 
$(\half,\half)$ or fundamental representations of ${\cal U}_{q \to 0}(su(2) \oplus 
su(2))$  and we get the results, reported in Table \ref{table-mitan}, where we have also written the observed anticodon for the mitochondria of animals taken from   \citep{Sprinzl}.

The purpose of this seminar  is double. At first we will  propose a mathematical approach, in the framework of the  ``crystal basis model" of the genetic code,  to determine which anticodon is chosen to translate the genetic information stored into the quadruplets and the doublets of codons \citep{SS12}. To succeed in this step, the idea is to require
the minimization of a suitable operator or function, mathematically expressed in terms of the quantities defined in the model, to explain why and which anticodon is used to   ``read"  more than a codon\footnote{We do not discuss here the chemical modified structure of the nucleotides, e.g see \citep{Agris}.}.
Then, we will show that this  scheme is well adapted to analyze and  model the evolution of the genetic code, with the two important steps, characterized by the Ancient Genetic Code later followed by the Early one, without forgetting the alternative primordial proposal of the Archetypal genetic Code \citep{SS13}.

\section{ PART 1: A Minimum Principle in Codon-Anticodon \\ Interaction}

The translational-transcription process from DNA to proteins is  a very complex process carried on in several steps. A key step is the translation from coding sequences of nucleotides in mRNA to the proteins chaines. In this process a role is played by the tRNA in which a triplet of nucleotides (anticodon) pairs to the triplet of nucleotides (codon) reading the genetic information.
Since there are 60 codons  (in mitochondrial code) specifying amino acids, the cell should contain 60 different tRNA molecules, each with a different anticodon in order to have a pairing codon anticodon following the usual Watson-Crick pattern, i.d. the pairing respectively  between the nucleotides C and G, and U and A. Actually, however, the number of observed anticodons is less than 60. This implies that an anticodon  may pair to  more than one codon. Already in the middle of the sixties,
 it was realized that the pairing anticodon-codon does not follow the standard rule and Crick \citep{Crick} proposed, on the basis of  the base-pair stereochemistry, the  ``wobble hypothesis".   According to this hypothesis a single tRNA type,  with a a specified anticodon, is able to recognize two or more codons in particular differing only in the third nucleotide, i.e only the first two nucleotides of a codon triplet  in mRNA have the standard precise pairing with the bases of the tRNA anticodon while the first nucleotide in the anticodon may pair to more than a nucleotide in the third position of the codon.  

This rule has been subsequently widely confirmed and extended, with  a better understanding of the chemical nucleotide modifications, for a review see \citep{Agris}. 
Since the years seventies the questions  were raised  \citep{Jukes}: how many anticodons do we need?  which anticodons do manifest? 

In order to explain which anticodon do manifest two main hypothesis have been advanced: \begin{enumerate}
\item The conventional wobble versatility hypothesis assumes that the the first position of anticodon should have G (U) to read for codon with Y (respectively R) in third position.
\item The codon adaptation hypothesis states that the first position of anticodon should pair the most abondant codon in the family of synonymous codons.
\end{enumerate}
For a comparison and discussion of the two hypothesis in fungal mitochondrial genomes  and for marine bivalve mitochondrial genomes, see \citep{CX} and \citep{YL}.

 In order to have a correct translation process between codons and amino-acids in the mitochondrial code we need a minimum number of 22 anticodons. In fact, in this code, the 20 amino-acids (a.a) are encoded by  2 sextets,
6 quadruplets and 12 doublets of codons. Considering a sextet as the sum of a quadruplet and a doublet, we need to dispose at least  of  22 anti-codons, of which 8 should  ``read" the quadruplets and 14 the doublets.  Indeed this seems to happen for the mitochondria of animals  \citep{Sprinzl,HJJR,WN,Nial,NikW}.  
The data seem to confirm the empirical  rule that the most used anticodons have as second and third nucleotide, respectively, the complementary  to the first and second nucleotide of the codons, while the first nucleotide is U for the anticodons pairing the quadruplets, G  and U for the anticodons pairing, respectively, the doublets ending with a pyrimidine and  with a purine, with exception of Met.

 \section{The  ``minimum" principle}
 
Given a  codon\footnote{In the paper we use the notation $N = C, A, G, U.; \; \;  R =   G, A. \;(purine)  ;\;\; Y =   C, U. \; (pyrimidine)$.} $XYZ$ ($X,Y,Z \in \{ C, A, G, U\}$) we conjecture that an anticodon    
 $\,X^aY^aZ^a$, where $\,Y^aZ^a=Y_cX_c$,  $N_c$ denoting  the nucleotide complementary to the nucleotide $N$ according to the Watson-Crick pairing rule\footnote{This property is observed to be verified in most, but not in all, the observed cases. To simplify we shall assume it.},
pairs to the codon $XYZ$, i.e. it is most used to ``read" the codon $XYZ$ if it minimizes the operator ${\cal T}$,  explicitly written in eq.(\ref{eq:T}) and computed between the 
 ``states", which can be read from Table \ref{table-mitan}, describing the codon and anticodon in the  ``crystal basis model". We write both codons (c) and anticodons (a) in $5" \to 3"$ direction. As an anticodon is antiparallel to codon, the 1st nucleotide (respectively the 3rd nucleotide) of the anticodon is paired to the 3rd (respectively the 1st) nucleotide of the codon, see Figure 1.
  
\be
{\cal T} = 8 c_H \,\vec{ J_H^c} \cdot  \vec{J_H^a} + 8  c_V \, \vec{J_V^c} \cdot \vec{ J_V ^a} \label{eq:T}
\ee
where:
\begin{itemize}

  \item $c_H.  c_V$ are constants depending on the  ``biological species" and weakly depending on the encoded a.a., as we  will later specify.
 
\item $J_H^c ,  J_V^c$  (resp. $J_H^a ,  J_V^a$) are the labels of  $U_{q \to 0}(su(2)_H \oplus su(2)_V)$  specifying the state \citep{FSS98} describing the codon $XYZ $ (resp. the anticodon $NY_cX_c$ pairing the codon  $XYZ$).

\item  $\vec{ J_{\alpha}^c} \cdot  \vec{J_{\alpha}^a}$ ($\alpha = H, V$) should be read as 
\be
\vec{ J_{\alpha}^c} \cdot  \vec{J_{\alpha}^a} =
\frac{1}{2}\left\{\left(\vec{J_{\alpha}^c}  \oplus \vec{J_{\alpha}}^a \right)^2 - ( \vec{J_{\alpha}^c})^2 -  ( \vec{J_{\alpha}^a})^2 \right\}
\ee
and   $\vec{J_{\alpha}^c}  \oplus  \vec{J_{\alpha}^a} \equiv \vec{J_{\alpha}^T}$  stands for the irreducible representation which the codon-anticodon state under consideration belongs to, the tensor product of   $\vec{J_{\alpha}^c}$ and $\vec{J_{\alpha}^a}$ being performed according to the rule of \citep{Kashi}, choosing the codon as first vector and the anticodon as second vector.  Note that 
$ \vec{J_{\alpha}}^2$ should be read as the Casimir operator whose eigenvalues are given by $
J_{\alpha}(J_{\alpha} +1 )$.
\end{itemize}

Let us  discuss in some detail how we compute the value of of the operator ${\cal T}$ defined in eq.(\ref{eq:T}) between the state of a codon $XZN$ and the state of an anticodon $WZ_cX_c$, i.e. $\; < XZN \, | \;{\cal T} \; |\,WZ_cX_c\,> $.  In the crystal basis model, there is a correspondence, see e.g. Table \ref{table-mitan},
\be
|XZN >  \; \ra |J_H^c, J_V^c;J_{H,3}^c, J_{V,3}^c > \;\;\;\;\,   |WZ_cX_c>  \; \ra |J_H^a, J_V^a;J_{H,3}^a, J_{V,3}^a >
\ee
We compute 
\be
<XZN|{\cal T}|WZ_cX_c >  \equiv \lambda 
\ee
where the value of $\lambda$ is given by the eigenvalue of  $\,{\cal T}$  on the state $|J_H^c, J_V^c;J_{H,3}^c, J_{V,3}^c > \otimes \,  |J_H^a, J_V^a;J_{H,3}^a, J_{V,3}^a > $, i.e.
we have
\bea
&& {\cal T} \, \left( |J_H^c, J_V^c;J_{H,3}^c, J_{V,3}^c > \otimes \,  |J_H^a, J_V^a;J_{H,3}^a, J_{V,3}^a >  \right) = \nonumber \\
&& \;\;\;\;\, \lambda \,   \left( |J_H^c, J_V^c;J_{H,3}^c, J_{V,3}^c > \otimes \, |J_H^a, J_V^a;J_{H,3}^a, J_{V,3}^a >  \right) 
\eea
For example the value  of ${\cal T}$  between the anticodon UUU and the codon AAC
is, using Table \ref{coef-d}:
\be
 < UUU| {\cal T}|AAC> =  -6 \, c_H +  18 \, c_V
 \ee
As we are interested in finding the composition of the 22  anticodons, minimun number to ensure a faithful translation,  we shall assume that the used anticodon for each quartet and each doublet is the one 
which minimizes the  averaged value of the operator  given in eq.(\ref{eq:T}), the average being performed over the 4 (2) codons  for quadruplets (doublets), see next  section.

\section{Structure of the minimum number of anticodons}

According to our conjecture on the existence of a minimum principle  we  determine,  for each quadruplet (q) and each doublet (d),  the anticodon which minimizes the averaged value  ${\cal T}_{av}$ of the operator ${\cal T}$ (see below).  We analyse separately the case of quadruplets and doublets.
 
\subsection{Quadruplets}

Let us give an example of what we mean by averaged value of ${\cal T}$. For example let us 
consider the anticodon  CAC for the a.a. Val,  we have to compute
\bea
&&  {\cal T}_{av}(CAC, Val) = \sum _{N} \; P^q_N \, <CAC|{\cal T}|GUN>  \nonumber \\
&& =  P^q_C  \, <CAC|{\cal T}|GUC> +  P^q_U  \, <CAC|{\cal T}|GUU>   \nonumber \\
&& + P^q_G  \,  <CAC|{\cal T}|GUG>+ P^q_A  \,   <CAC|{\cal T}|GUA>  \nonumber  \\
&& = 2 (P^q_C + P^q_U + P^q_G +P^q_A) \, c_H + (6 P^q_C + 6 P^q_U + 2 P^q_G + 2P^q_A) \, c_V\nonumber \\
&& = 2 \, c_H  + [6 P^q_Y + 2(1- P^q_Y)]\, c_V
\eea
In the computation we have to take into account the codon usage frequency or relative percentage of the appearance of each codon in the quadruplet and we have denoted with $P^q_N$ the codon usage frequency for codon ending with N.
Really  we need to introduce the following four positive frequencies
 $P_Y^q$ $P_R^q$, $P_S^q$,$P_W^q$, with the normalization condition:
\be
P_Y^q + P_R^q = P_S^q + P_W^q = 1
\ee
where, respectively, $P_Y^q$, $P_R^q$, $P_S^q$ and $P_W^q$ denote the relative usage frequency of the codons ending  with nucleotides C,U (pyrimidine),  G,A (purine),  C,G and  U,A.
From Table  \ref{coef-q} we can compute the value which we report in Table \ref{avT-q}.
 
  \subsection{Doublets}

In the computation we have to take into account the codon usage frequency in the doublet. Now we need to introduce the following four positive frequencies
 $P_C^d$, $P_U^d$, $P_G^d$, $P_A^d$, with the normalization condition
\be
P_C^d + P_U^d = P_G^d + P_A^d = 1
\ee
 As example let us compute the averaged value of ${\cal T}$ for Asp. 
Let us consider the anticodon  CUC  we have to compute
\bea
&&{\cal T}_{av}(CUC, Asp) = \sum _{Y} \;  P_Y^d\,  <CUC|{\cal T}|GAY>  = P_C^d\, <CUC|{\cal T}|GAC> + P_U^d\,  <CUC|{\cal T}|GAU>    \nonumber  \\
&& = 2 \, c_H +18 \, c_V\
\eea
From Table  \ref{coef-d},  we can compute the values, which we report in Table \ref{avT-d2}. 

 Let us remark that:
 \begin{itemize}
\item  for all a.a. the contribution of $su_V(2)$ verifies the same property than for the quadruplets and, moreover, is not depending on the codon usage;
\item for 4  a.a.  the contribution of $su_V(2)$ is the same for all anticodon.
\end{itemize}
 
 From the above remarks we easily realize that the case of doublets is more complicated than the one of the quadruplets. In some sense the contribution of $su_V(2)$  plays a role only in establishing the  most preferred anticodon. Moreover,  as we do not want a priori to exclude any anticodon, we have to face the possibility that an anticodon can be chosen to read for more than one doublet. In order to avoid this problem, in contradiction with the requirement of a faithful translation process, we make the following choice:
 \begin{enumerate}
 \item the sign of the constant $c_H$ for the doublets ending with a purine is the opposite of the sign of
 the doublets ending with a pyrimidine with the same dinucleotide (if it does exist)\footnote{We call dinucleotide the first two nucleotids of the codon.}. 
 \item the sign of  $c_H$ for the 8 weak dinucleotides encoding doublets is positive for the following 4 doublets UUY, UAY, AUY, AAY and negative for the remaining 4, i.e. CAY, UGY, AGY, GAY.
 \end{enumerate} 
  and fix the following procedure, while considering doublets with the same dinucleotide:

 \begin{enumerate}
\item first we select, among the four possible anticodons, the one giving the lowest
value for  ${\cal T}$ averaged on the two codons of each doublet and assign this anticodon to
the corresponding doublet.

\item then the anticodon reading the second doublet is chosen between the two ones
containing as a first nucleotide a purine, resp. a pyrimidine, if the first
nucleotide of the anticodon already determined for the first doublet is a
pyrimidine (resp. a purine).
\end{enumerate}  
 
As an illustration, we take the case of the Cys and Trp amino acids. The anticodon
GCA can minimize both of them, but more Cys (due to the value $-6$ for the $c_V$ coefficient) than Trp
(with value $2$ for the same  $c_V$ coefficient). Thus GCA will be taken as the anticodon
relative to Cys, while the choice for the Trp anticodon will be made between UCA and
CCA, that is the two candidates with a pyrimidine as a first nucleotide, the
anticodon GCA starting with a purine.

Let us remark that, even if the above assumptions seem rather ad hoc,  indeed a general symmetric pattern shows up: for half of a.a. $c_H$ is positive and for the other half is negative;  the first set of 4 dinucleotides involves only  `weak" nucleotides, the second one a  ``strong" nucleotide; the dinucleotides XY and YX correspond to the same sign.

 \subsection{Discussion}

  For all quadruplets, from Table  \ref{coef-q}, we remark that for $c_H > 0$ and  $c_V < 0$ the  anticodon  minimizing  the average value of ${\cal T}$ has the composition $ UX_cY_c$. For Leu, Val and Thr a very weak condition for codon usage frequency has to be satisfied, i.e for the first two a.a. $  P^q_S  > 0,25$ and for the last one $  P^q_Y  > 0,125$\footnote{ The constraint on the codon usage frequency  can be released, imposing a suitable condition between $c_H$ and  $c_V $.}.
   The results are in agreement with the observed anticodons, see \citep{Sprinzl} and Table \ref{table-mitan}.
 
 For doublets, we remark that, with the choice of sign of  $c_H $  above specified and  $c_V > 0$ for all a.a., the  anticodons  minimizing  the average value of ${\cal T}$  are  in agreement with the observed anticodon, see \citep{Sprinzl} and Table \ref{table-mitan}.
 We summarize in Table \ref{min-d} the results for the doublets.

   Let us remark that we find that for Met the anticodon is not UAU, as it should be expected from the empirical rule  above quoted, but CAU which seems in agrement with the data, see \citep{Sprinzl}.

We have found that the anticodons minimizing the conjectured operator ${\cal T}$ given in eq.(\ref{eq:T}), averaged  over the concerned multiplets, are in very good agreement, the results depending only on the signs of the two coupling constants, with the observed ones, even if we have made comparison with a limited database.

In the present approach we have faced the problem to find the structure of the mimimum set of anticodons and, then, we have used a very simple form for the operator ${\cal T}$.  
 We have not at all  discussed the possible appearance of any  other anticodon, which should require a more quantitative discussion.
For such analysis, as well as
for the eukaryotic code, the situation may be different and more than an anticodon may pair to  a quartet.  For this future aim we have here reported Tables \ref{coef-q} and  \ref{coef-d}. 

The pattern, which in the general case may show up, is undoubtedly more complicated, depending on the biological species and on the concerned biosynthesis process, but it is natural to argue that the usage of anticodons  exhibits the general feature to assure an ``efficient"  translation process by  a number of anticodons, minimum with respect to the involved constraints.
 A more refined and quantitative analysis, which should require more data,  depends on  the value of these constants.
Very likely, it might happen that the assumption of the  ``universal" feature of $c_H$ and $c_V$ should be released and that  the expression of the operator eq.(\ref{eq:T}) should be modified, for example by adding a term of  ``spin-spin" interaction of the type
\be
4 g_H \, J_{H,3}^c J_{H,3}^a +   4 g_V \, J_{V,3}^c J_{V,3}^a 
\ee
where the values of  $ J_{H,3}$ and $ J_{V,3} $,  both for codons and anticodons, can be read out  from their nucleotide composition, see  Table \ref{table-mitan}.
However the pattern which shows up in Tables \ref{coef-q} and \ref{coef-d}, with the values of the coefficients  equal in pairs, strongly suggests that the minimum number of anticodons should be 32 (3 for the sextets, 2 for quadruplets and triplet and 1 for doublets and singlets).

\section{PART 2:  Codon-antocodon Interaction and the Genetic Code Evolution}

For long time the genetic code was thought to be immutable. Now it is commonly believed that the genetic code has undergone (and is undergoing) an evolutionary process, see \citep{Knight01} for a review, even if there is not a unique theory of the evolution.

It has been proposed that the evolution of the genetic code is based on the ``codon capture theory'', see \citep{Ohama} for a recent review. In this scheme, the number of the encoded amino-acids (a.a.) is kept constant and equal to 20 and the coding codons change in the evolution, a key role in this process being played by the anticodon. Inside the scheme of the fixed, from the very beginning,  total number of a.a., it has been proposed a mechanism for the reassignment of the codons, alternative to the codon capture theory, called ``ambiguous intermediate scenario'' \citep{SY96}.  The main difference between the two schemes is that  in the first  one   ``codon disappearance''  by mutation pressure  is assumed while in the latter an anticodon can read more than one codon.  A comparison between that two schemes based on  a physical model is presented in \citep{YN11}.

One can distinguish two important steps in the genetic code evolution, characterized by the Ancient Genetic Code later followed by the Early one. An alternative primordial code has been proposed by \citep{Jukes83} and called  ``Archetypal Genetic Code��, in which only sixteen anticodons are involved.

Using again the minimum principle approach developed in Part 1, with  the operator ${\cal T}$ as already defined, we propose to analyze  the evolution of the genetic code in this spirit. More precisely,  examining each of the three above mentioned genetic codes from the codon-anticodon interaction, itself evolving in time, we propose a mathematical scenario exhibiting the successive passages from the Ancient to the Archetypal code and then to the Early one, thus conciliating these three codes.

Let us remark that in our approach we have mainly followed some simple assumptions, which we still  keep hereafter:
\begin{itemize}
\item we have assumed a simple YES or NO model, i.e. the only anticodon  chosen is the one which minimizes the interaction. Indeed one should consider that a codon is chosen with a probability related to the value of the interaction,
\item in the choice of the solutions we have independently minimized each term depending on $c_H$  and $c_V$ to prevent  solutions depending on a fine tuning between the two constants. The values of the constants, even if not explicitly indicated in order not to make heavier the  notation, depend from the multiplet encoding a determined a.a.\footnote{We neglect the dependence on the particular codon in the multiplet.},
\item we have assumed the codon usage frequency as a fixed external data. This is not correct as the  usage of a particular codon depends on the total  G+C content and on the presence and abundance of the anticodon(s) which can match with. Indeed   the frequency of the codon and the frequency of the anticodon are related by  a kind of ``bootstrap'' relation, i.e. by a self-consistent process ruled by the reciprocal effects codon-anticodon.
\end{itemize}

\section{The Evolution of the Genetic Code}

In this Section we discuss in our scheme the evolution from the Ancient Code to the Early one, very similar to the Mitochondrial one.

\subsection{The Ancient Genetic Code}

In this code it is assumed \citep{Ohama} that any a.a. is encoded by only one codon, that is 20 codons are assigned to codify, one is assigned to the stop function and the other 43 are unassigned. Which codon is used to encode? We make the following assumptions:
\begin{enumerate}
\item the Watson-Crick pairing mechanism for nucleotides holds in the  codon-anticodon coupling,
\item the coding codon belongs to the multiplet encoding, in the present Standard  Genetic Code,  the concerned a.a.. For the sextet, which we consider as the sum of a quartet and a doublet, we consider only the quartet sub-part,
\item the selected codon-anticodon couple is the one whose mutual interaction,  as described in Part1 and according to the assumption 1),   is minimum,
\item the coupling constants have the same sign for all  ``strong'' and weak''  dinucleotides. We call dinucleotide the first two nucleotides of a codon. We call a dinucleotide XZ  ``strong'' if in the Standard Code the quartet XZN (N $\in$ C,U,G,A)  encodes the same a.a., otherwise we call it ``weak''.
\end{enumerate}

The coupling constants  $c_H$ and  $c_V$ are selected as it follows:
\begin{itemize}
 \item $c_V < 0$ for the strong dinucleotides  while its sign is underdetermined for the weak ones,
 \item  $c_H < 0$  for the strong dinucleotidfes 
  and  $c_H > 0$    for the weak dinucleotides.
 \end{itemize}  
 
  We do not discuss here the  modified chemical structure of the nucleotides, e.g.  see \citep{Agris}, and use the same symbol and the same algebraic mathematical structure for nucleotides  inside a codon or an anticodon. 
  We report in Table \ref{codacod-agc} the unique couple codon-anticodon minimizing the interaction. 
 We have written in bold the differences with Table 2 of  \citep{Ohama}, namely for Leu, Val, Cys and  Trp. \footnote{If  $c_H > 0$ for CU. and GU. then we get for Leu and Val the same result of \citep{Ohama}.  Let us also note that these dinucleotides in our model  have a peculiar structure as they belong to the irreducible representation (0, 1) \citep{FSS01}.}.
  Let us remark that,  by  merely considering the hydrogen bond property between the nucleotides C and G, there is no reason why it  happens that the chosen codon has, for example, the structure XZC and the anticodon,   $GZ_cX_c$   and not the contrary, that is codon XZG and an anticodon  $CZ_cX_c$, i.e. why the nucleotide C is preferred in the codon and the nucleotide G in the anticodon. Our model  predicts the nature of the couples codon-anticodon in agreement with the generally  accepted pattern for most of the a.a..
  
  The simple  Ancient Code  is
not viable because too sensitive to any mutation  of the codon or translation errors which would imply either the change of the a.a. in the protein chain or the breakdown of the translation process. Therefore,  by means of a mechanism,  not yet completely clear, of interplay between mRNA and tRNA,  through the use of synonymous codons, each a.a. is encoded by multiplet, going from doublet to sextet,  not taking into account the singlet and triplet which are present in the Standard Code. 

\subsection{The Archetypal Genetic Code}

The Archetypal Genetic Code has been proposed \citep{Jukes83} as  an ancestor of the present genetic codes. In our scheme it appears as an evolution of the Ancient Code.  The lack of effectiveness of the Watson-Crick pairing and the appearance of the  ``wobble'' mechanism expands the encoding codon from one to a quartet, e.g. from XZC to XZN.

It follows that, in the computation, we have now to take into account the codon usage frequency or relative percentage of the appearance of each synonymous codon in the multiplet  and we have denoted with $P_N$ the codon usage frequency or more precisely the usage probability for codon ending with N, omitting the dependence on the first two nucleotides,  with the normalization, for a quartet,
\be
\sum_{N} \, P_N = 1
\ee
It is also convenient to introduce the following four probabilities
 $P_Y$, $P_R$, $P_S$,$P_W$, with the normalization condition:
\be
P_Y+ P_R = P_S + P_W = 1
\ee
where, respectively, $P_Y$, $P_R$, $P_S$ and $P_W$ denote the  usage frequency of the codons ending,  respectively, with nucleotides C,U (pyrimidine),  G,A (purine),  C,G and  U,A.

Indeed, in order to find which anticodon matches with four codons we have to find for the minimum of the operator $ {\cal T}$ averaged over the multiplet, i.e. 
\bea
&&  {\cal T}_{av}(N^aZ^a_cX^a_c, \,X^cZ^cN^c)= \sum _{N} \; P_N \, <N^aZ^a_cX^a_c|{\cal T}|X^cZ^cN^c>  \nonumber \\
&& =  P_C  \, <N^aZ^a_cX^a_c|{\cal T}|X^cZ^cC> +  \, P_U  \, <N^aZ^a_cX^a_c|{\cal T}|X^cZ^cU>   \nonumber \\
&& + P_G  \,  <N^aZ^a_cX^a_c|{\cal T}|X^cZ^cG>+ \, P_A  \,   <N^aZ^a_cX^a_c|{\cal T}|X^cZ^cA>  
\eea
where the value of $ <N^aZ^a_cX^a_c|{\cal T}|X^cZ^cN^c>$ can be read from Tables 3 and 4.

For the quadruples  XZN which correspond to encoding quartets  in the Standard Genetic Code
the Table 3 of  \citep{Ohama}  is equal to Table 2  with  $c_H > 0$ and  $c_V < 0$.
The quadruple UAN is a particular one since it contains the Stop codons.
For  the doublet UAY, encoding Tyr,\footnote{For this doublet the operator $ {\cal T}$ has to averaged over UAC and UAU.}  the selected anticodon is  GUA for  $c_H > 0$ and  $c_V > 0$.
 For the other quadruples (i.e. UUN, AUN, UGN, CAN, AAN, AGN and GAN) we cannot immediately make a comparison between our results and the Table 3 of  \citep{Ohama} as our calculations were done there for the Mitochondrial Genetic  Code. 
 
 Carrying on  the calculations along the same line of reasoning  done in Part1, we get the  results reported in Table \ref{sign-arco} of the present paper, which are the same as those reported in Table 3 of  \citep{Ohama}.
  
\subsection{The Early Genetic Code}

The Archetypal Code suffers of some serious drawbacks. Indeed it can be considered either to encode only 15 a.a. or to have not defined encoding correspondence for several a.a.. To encode definitively and unambiguously all a.a. a further step is needed. A natural way to proceed is the splitting of some (weak) quadruplets into doublets, leading to the Early Genetic Code. 
Let us note that for doublets we have now $\sum_{N \in R} \, P_N = 1$ \, or \,  $\sum_{N \in Y} \, P_N = 1$. In Table \ref{sign-eco} we report our results  where we have noted in bold the difference with  the Table 4 of  \citep{Ohama}.
 
 The entries of  Table \ref{sign-eco} of the present paper are the same as those reported in Table 2, confirming that the Early Code is not much different from the present Mitochondrial Code. 

\section{Discussion}

Let us discuss  and comment our results. The pattern of the Ancient Code, as reported in Table 2 of \citep{Ohama}, can be summarized by saying that in this primordial code the a.a.,
which would be encoded by a doublet of the type XZY are encoded by a codon ending with a C.
Similarly the a.a. which would be encoded by a doublet of the type XZR are encoded
by a codon ending with a G.
 As already remarked by  merely considering the hydrogen bond between the nucleotides, there is no reason  for such a pattern, while our scheme  predicts, as a consequence of the structure of the model and of the hypothesis of the minimization requirement, obtained only with a constraint on the the sign of $c_H$ and  $c_V$, the observed configuration for 6 of the 8 a.a. encoded, in the Standard Code, by at least a quartet and for 10 of the 12 a.a. encoded by a doublet. Moreover the pattern of the model is surprisingly coherent.
  
Indeed, in the Ancient Genetic Code, the sign of $c_V$ for the weak dinucleotides is undetermined,  i.e. the minimization does not depend on the sign of $c_V$. In our model, this means that there is not distinction between C (U) and G  (A). This is coherent since at this  stage there is not yet a distinction between the doublet XZR and XZY. On the contrary for strong dinucleotides  for which the role of  XZR and XZY is the same up to the  Standard Genetic Code, the sign is fixed and it does not change during the evolution.
For strong dinucleotides  and almost half of the weak ones\footnote{Let us remark that the sign of $c_H$ does not change for the weak dinucleotides which  has the value of $J_{3,H}= 0$. } there is a change in $c_H$  just when  the codon degeneracy appears, that is going from the Ancient to the Archetypal code, and the `` wobble mechanism ''  \citep{Crick} is called in. For all weak  dinucleotides, the sign  of $c_V$  is  now determined and  there is a further change in the sign of $c_H$  and of $c_V$  when the  correspondence between doublets and a.a. is fixed.  

Let us make a few comments on the change of the sign of  $c_H$, see Table  \ref{sign-evol} and Figure \ref{cH.pdf}.  The present formulation of  the model is  too general to allow a study of the change of the constant, i.e. if  $c_H$ has a smooth  or  a phase transition like behavior through  the point   $c_H =0$. We remark that at this value the interaction is  only ruled by the value of $c_V$ and that, from the Tables 3 and 4, for each codon there are at least two anticodons with the same value of  ${\cal T}$ and viceversa. This degeneracy can be removed by further terms of the interaction, not yet taken into account, but it can be also read as the ``codon disappearance''  before a new readjustment of the code.
 
 Looking at the data of codon usage frequencies  for the Mitochondrial Code from GenBank database \citep{data} for 30 species with high statistics, we remark that the presently less used (in the average)  codons,  for the a.a. encoded by doublets, are  those with last nucleotide G or U, while the most used are those with last nucleotide A or C.  So it is natural to ask the question:  why most of  the ancestral codons encoding a.a. in the Ancient Code, see Table  \ref{codacod-agc}, are now  repressed ? Naively, one should expect that the ancestral codon  should be the most used one.  Looking at Table \ref{sign-evol}  we see that in our model the sign of $c_H$ for a.a. encoded by XZY in the Early Code is the same than the one in the Ancient Code, while  for a.a. encoded by XZR is the opposite.   Cys  is an exception,  but in this case the anticodon  is different in the two codes.
 This kind of argument cannot be immediately applied to a.a. encoded by quartets because in most cases the anticodon in the Archetypal,  Early or Mitochondrial Code is not the same as the one appearing in the Ancient Code and, moreover, there is an important effect due to the averaging over four codons. 
 
  Analogous analysis of the codon usage frequencies for species following the Standard Code confirms generally such a pattern,  but the presence of anticodons in the Standard Code is more complicated, so we do not want to refer to these data.

Moreover, in our model naturally the anticodon with first nucleotide A does never appear, in good agreement with the observed data.

In our model we can express the evolution of the genetic code through the following pattern of the codon-anticodon interaction as 
\bea
&&  <XZN|{\cal T}|N^aZ^a_cX^a_c>  = <XZN| \left( 8 c_H \, \vec{ J_H^c} \cdot  \vec{J_H^a} + 8  c_V \, \vec{J_V^c} \cdot \vec{ J_V ^a}\right) \delta_{M^a,N^a_c}|M^aZ^a_cX^a_c> \nonumber \\
&&  \;\;\;\, \;\;\;\,\;\;\;\, \;\;\;\, \;\;\;\,\Lra_{Evolution} \;\;\;  <XZN| 8 c_H \,\vec{ J_H^c} \cdot  \vec{J_H^a} + 8  c_V \, \vec{J_V^c} \cdot \vec{ J_V ^a}|M^aZ^a_cX^a_c>  \label{eq:iaa}
\eea
Of course we have to assume that the constants $c_H$ and $c_V$ depend on the  ``time''. At this stage only the change  of the sign in the coupling constants has been considered.
In the first row of eq.(\ref{eq:iaa}), the presence of the delta di Kronecher $\delta_{M^a,N^a_c}$ enforces the Watson-Crick coupling mechanism implying $M^a = N^a_c$, while in the second row $M^a$ can be any nucleotide and the selection is implemented by the value of the operator ${\cal T}$, computed between the concerned states and, eventually, averaged over the multiplet taking into account the codon usage probabilities.  As example of  typical behavior of the constant $c_H$ for  weak dinucleotides, we consider the case of the AA dinucleotide, see Table \ref{sign-evol}
\bea
c_H^{AAN}  > 0 \; \Lra \;  c_H^{AAN}  <  0 \;  \Lra \;  \left\{ \begin{array}{c} \;  c_H^{AAY}  > 0 \\ \\
 \; c_H^{AAR}  < 0  \end{array} \right.
\eea
The change of the sign in the coupling constants is a mathematical description to frame the modification of the  interaction codon-anticodon due to the change of the molecular structure of the nucleotides in the anticodons and of the (non local) structure of the tRNA.

\section{Conclusions}

At first,  we recall that in the framework of this model,  a   ``reading''  (or ``ribosome'') 
operator  ${\cal R}$ has been derived which provides the correspondence between amino-acids and codons for any known genetic code \citep{FSS01}. In other words, two codons have the same
eigenvalue under  ${\cal R}$  if and only if they are associated to the same 
amino-acid. One can note that the passage from one
genetic code to another one is insured by the presence  - or absence - in the operator  ${\cal R}$  of some term  precisely constructed with algebraic quantities of the type as the ones introduced in eq.(\ref{eq:T}). Thus the construction of this ``reading''  operator for the different genetic codes can - a posteriori -  be 
considered as an algebraic attempt to study the evolution from one to another to-day existing codes. 

Let  us  emphasize that the fact that the crystal basis model is able to explain, in a  relatively simple way,  the observed anticodon-codon pairing which has its roots on the stereochemical properties of nucleotides \citep{LimCur} strongly suggests that our modelisation is able to incorporate some crucial features of the complex physico-chemical structure of the genetic code. Incidentally let us remark that the model  explains the symmetry codon anticodon remarked in \citep{WN}.  Let us stress that our modelisation has
a very peculiar feature which makes it very different from the standard 4-letter alphabet, used to identify the nucleotides, as well as with the usual modelisation of nucleotide chain as spin chain.  Indeed the identification of the nucleotides with the fundamental irrep. of ${\cal U}_{q}(su(2)_H \oplus su(2)_V)$ introduces a sort of  double ``bio-spin", which allows the description of any ordered sequence of $n$ nucleotides as as state of an irrep. and allows to describe interactions using the standard powerful mathematical language used in physical spin models.
Let us conclude this contribution by one final remark: among the important questions which deserve to be considered in the context of
evolution of the genetic code, the adaptability of the code with the increasing
complexity of the organisms  is a crucial one. It will be worthwhile to see to what
extend our methods can be used for such a problem in the light of the very
interesting approach considered in \citep{IA07}.
In \citep{S} a mathematical model, always in the framework of the    ``crystal basis model'' of the genetic code, has been presented in which the main features (numbers of encoded a.a., dimensions and structure of synonymous codon multiplet)  are obtained, requiring stability of the genetic code against mutations modeled by suitable  operators.
In conclusion our scheme appears rather well adapted  to reproduce a mathematical model able to reproduce the features of the codon capture theory as well as to provide a mathematical framework for a more quantitative and detailed description of the theory. 

\vspace{1cm} 

{\bf Ackmowledgment} - Paul Sorba would like to express his deepest gratitude to Professor Branko Dragovich for his invitation to this very interesting conference.

\newpage

\begin{table}[htbp]
\footnotesize
\begin{center}
\begin{tabular}{|cc|cc|rr|c||cc|cc|rr|c||}
\hline
codon & a.a. & $J_{H}$ & $J_{V}$ & $J_{3,H}$ & $J_{3,V}$& anticodon & codon & 
a.a. &   $J_{H}$ & $J_{V}$ & $J_{3,H}$ & $J_{3,V}$ & anticodon\\
\hline
\Big. CCC & P & $\Half$ & $\Half$ & $\Half$ & $\Half$ & & UCC & S & 
$\Half$ & $\Half$ & $\half$ & $\Half$ & \\
\Big. CCU & P & $(\half$ & $\Half)^1$ & $\half$ & $\Half$ &  
&UCU & 
S & $(\half$ & $\Half)^1$ & $-\half$ & $\Half$&  \\
\Big. CCG & P & $(\Half$ & $\half)^1$ & $\Half$ & $\half$ &{\LARGE \color{red} \bf UGG} & UCG & 
S & $(\Half$ & $\half)^1$ & $\half$ & $\half$  & {\LARGE \color{red} \bf   \bf UGA} \\
\Big. CCA & P & $(\half$ & $\half)^1$ & $\half$ & $\half$  & &UCA & 
S & $(\half$ & $\half)^1$ & $-\half$ & $\half$ & \\[1mm]
\hline
\Big. CUC & L & $(\half$ & $\Half)^2$ & $\half$ & $\Half$ & & UUC & 
F & $\Half$ & $\Half$ & $-\half$ & $\Half$ & \\
\Big. CUU & L & $(\half$ & $\Half)^2$ & $-\half$ & $\Half$ & 
& UUU & 
F & $\Half$ & $\Half$ & $-\Half$ & $\Half$ &  {\LARGE \color{blue} \sl GAA}\\
\Big. CUG & L & $(\half$ & $\half)^3$ & $\half$ & $\half$ & {\LARGE \color{red} \bf UAG} & UUG & 
L & $(\Half$ & $\half)^1$ & $-\half$ & $\half$ &   \\
\Big. CUA & L & $(\half$ & $\half)^3$ & $-\half$ & $\half$ & &UUA & 
L & $(\Half$ & $\half)^1$ & $-\Half$ & $\half$ & {\LARGE \color{blue} \sl UAA} \\[1mm]
\hline
\Big. CGC & R & $(\Half$ & $\half)^2$ & $\Half$ & $\half$ & &UGC & 
C & $(\Half$ & $\half)^2$ & $\half$ & $\half$& \\
\Big. CGU & R & $(\half$ & $\half)^2$ & $\half$ & $\half$ &   
& UGU & 
C & 
$(\half$ & $\half)^2$ & $-\half$ & $\half$  & {\LARGE \color{blue} \sl  GCA}  \\
\Big. CGG & R & $(\Half$ & $\half)^2$ & $\Half$ & $-\half$ &  {\LARGE \color{red} \bf UCG} & UGG & 
W & $(\Half$ & $\half)^2$ & $\half$ & $-\half$ &\\
\Big. CGA & R & $(\half$ & $\half)^2$ & $\half$ & $-\half$ & &UGA & 
W & $(\half$ & $\half)^2$ & $-\half$ & $-\half$ & {\LARGE \color{blue} \sl  UCA} \\[1mm]
\hline
\Big. CAC & H & $(\half$ & $\half)^4$ & $\half$ & $\half$ &  &UAC & 
Y & $(\Half$ & $\half)^2$ & $-\half$ & $\half$ &  \\
\Big. CAU & H & $(\half$ & $\half)^4$ & $-\half$ & $\half$ & {\LARGE \color{blue} \sl GUG} & UAU & 
Y & $(\Half$ & $\half)^2$ & $-\Half$ & $\half$ &  {\LARGE \color{blue} \sl GUA} \\
\Big. CAG & Q & $(\half$ & $\half)^4$ & $\half$ & $-\half$ &&UAG & 
 {\bf Te}r & $(\Half$ & $\half)^2$ & $-\half$ & $-\half$ & ----- \\
\Big. CAA & Q & $(\half$ & $\half)^4$ & $-\half$ & $-\half$ & {\LARGE \color{blue} \sl UUG}  &UAA & 
 {\bf Ter} & $(\Half$ & $\half)^2$ & $-\Half$ & $-\half$& ----- \\[1mm]
\hline
\Big. GCC & A & $\Half$ & $\Half$ & $\Half$ & $\half$ & & ACC & T & 
$\Half$ & $\Half$ & $\half$ & $\half$ &  \\
\Big. GCU & A & $(\half$ & $\Half)^1$ & $\half$ & $\half$ && ACU & 
T & $(\half$ & $\Half)^1$ & $-\half$ & $\half$ & \\
\Big. GCG & A & $(\Half$ & $\half)^1$ & $\Half$ & $-\half$ &  {\LARGE \color{red} \bf UGC} & ACG & 
T & $(\Half$ & $\half)^1$ & $\half$ & $-\half$& {\LARGE \color{red} \bf UGU} \\
\Big. GCA & A & $(\half$ & $\half)^1$ & $\half$ & $-\half$ & &ACA & 
T & $(\half$ & $\half)^1$ & $-\half$ & $-\half$ &\\[1mm]
\hline
\Big. GUC & V & $(\half$ & $\Half)^2$ & $\half$ & $\half$ &  & AUC & 
I & $\Half$ & $\Half$ & $-\half$ & $\half$& \\
\Big. GUU & V & $(\half$ & $\Half)^2$ & $-\half$ & $\half$ & & AUU & 
I & $\Half$ & $\Half$ & $-\Half$ & $\half$&   {\LARGE \color{blue} \sl GAU} \\
\Big. GUG & V & $(\half$ & $\half)^3$ & $\half$ & $-\half$ & {\LARGE \color{red} \bf UAC}& AUG & 
M & $(\Half$ & $\half)^1$ & $-\half$ & $-\half$&  \\
\Big. GUA & V & $(\half$ & $\half)^3$ & $-\half$ & $-\half$ & &
AUA &  M & $(\Half$ & $\half)^1$ & $-\Half$ & $-\half$ &  {\LARGE \color{blue} \sl CAU}\\[1mm]
\hline
\Big. GGC & G & $\Half$ & $\Half$ & $\Half$ & $-\half$ & & AGC & S 
& $\Half$ & $\Half$ & $\half$ & $-\half$ & \\
\Big. GGU & G & $(\half$ & $\Half)^1$ & $\half$ & $-\half$ &&AGU & 
S & $(\half$ & $\Half)^1$ & $-\half$ & $-\half$ &  {\LARGE \color{blue} \sl GCU}\\
\Big. GGG & G & $\Half$ & $\Half$ & $\Half$ & $-\Half$ & {\LARGE \color{red} \bf UCC} &AGG & 
{\bf Ter} & $\Half$ & $\Half$ & $\half$ & $-\Half$&----- \\
\Big. GGA & G & $(\half$ & $\Half)^1$ & $\half$ & $-\Half$ & 
&AGA & {\bf Ter} & $(\half$ & $\Half)^1$ & $-\half$ & $-\Half$ & ----- \\[1mm]
\hline
\Big. GAC & D & $(\half$ & $\Half)^2$ & $\half$ & $-\half$ && AAC & 
N & $\Half$ & $\Half$ & $-\half$ & $-\half$  & \\
\Big. GAU & D & $(\half$ & $\Half)^2$ & $-\half$ & $-\half$ & {\LARGE \color{blue} \sl GUC} & AAU & 
N & $\Half$ & $\Half$ & $-\Half$ & $-\half$ & {\LARGE \color{blue} \sl  GUU} \\
\Big. GAG & E & $(\half$ & $\Half)^2$ & $\half$ & $-\Half$ &   & AAG & 
K & $\Half$ & $\Half$ & $-\half$ & $-\Half$ & \\
\Big. GAA & E & $(\half$ & $\Half)^2$ & $-\half$ & $-\Half$ &{\LARGE \color{blue} \sl UUC} & AAA & 
K & $\Half$ & $\Half$ & $-\Half$ & $-\Half$& {\LARGE \color{blue} \sl UUU} \\[1mm]
\hline
\end{tabular}
\caption{
The vertebral mitochondrial code. The upper label denotes 
different irreducible representations. 
 We list the most used anticodons for mitochondria of animals, see \citep{Sprinzl}. In bold-red (italic-blue) the anticodons reading quadruplets (resp. doublets).}
\label{table-mitan}
\end{center}
\end{table}

\clearpage
\newpage
\begin{table}[htbp]
\begin{tabular}{|c|c||c|c||c|c||c|c||c|c||} \hline 
a.a & codon &$ K_{H,C}$ & $ K_{V,C}$   &$ K_{H,U}$ & $ K_{V,U}$ &$ K_{H,G}$ & $ K_{V,G}$&   $ K_{H,A}$ & $ K_{V,A}$  \\
\hline \hline
Pro & CCC & 18 & -10& -6 & -10& 18 & -30& -6 & -30 \\
& CCU & 6 & -10& -10 & -10& 6 & -30& -10 & -30 \\
& CCG & 18 & -6& -6 & -6& 18 & -10& -6 & -10 \\
& CCA & 6 & -6 & -10 & -6 & 6 & -10 & -10 & -10 \\
\hline 
Leu & CUC & 2 & -10& -10 & -10& 2 & -30& -10 & -30 \\
& CUU & 2 & -10& 6 & -10& 2 & -30& 6 & -30 \\
& CUG & 2 & -6& -10 & -6& 2 & -10& -10 & -10 \\
& CUA & 2 & -6 & 6 & -6 & 2 & -10& 6 & -10 \\
\hline
Arg & CGC & 18 & 2 & -6 & 2 & 18 & -6 & -6 & -6 \\
& CGU & 6 & 2 & -10 & 2 & 6 & -6 & -10 & -6 \\
& CGG & 18 & 2 & -6 & 2 & 18 & 2 & -6 & 2 \\
& CGA & 6 & 2 & -10 & 2 & 6 & 2 & -10 & 2 \\
\hline
Ala  & GCC & 18 & 6 & -6 & 6 & 18 & -22 & -6 & -22 \\
& GCU & 6 & 6 & -10 & 6 & 6 & -22 & -10 & -22 \\
& GCG & 18 & 2 & -6 & 2 & 18 & 6 & -6 & 6 \\
& GCA & 6 & 2 & -10 & 2 & 6 & 6 & -10  & 6 \\
\hline
Gly  & GGC & 18 & 18 & -6 & 18 & 18 & -6 & -6 & -6 \\
& GGU & 6 & 18 & -10 & 18 & 6 & -6 & -10 & -6 \\
& GGG & 18 & 18 & -6 & 18 & 18 & 18 & -6 & 18 \\
& GGA & 6 & 18 & -10 & 18 & 6 & 18 & -10  & 18 \\
\hline 
Val & GUC & 2 & 6 & -10 & 6 & 2 & -22& -10 & -22 \\
& GUU & 2 & 6 & 6 & 6 & 2 & -22& 6 & -22 \\
& GUG & 2 & 2 & -10 & 2 & 2 & 6 & -10 & 6 \\
& GUA & 2 & 2 & 6 & 2 & 2 & 6 & 6 & 6 \\
\hline
Ser & UCC & 6 & -10& -10 & -10& 6 & -30& -10 & -30 \\
& UCU & 2 & -10& 2 & -10& 2 & -30& 2 & -30 \\
& UCG & 6 & -6& -10 & -6& 6 & -10& -10 & -10 \\
& UCA & 2 & -6 & 2 & -6 & 2 & -10 & 2 & -10 \\
\hline
Thr  & ACC & 6 & 6 & -10 & 6 & 6 & -22 & -10 & -22 \\
& ACU & 2 & 6 & 2 & 6 & 2 & -22 & 2 & -22 \\
& ACG & 6 & 2 & -10 & 2 & 6 & 6 & -10 & 6 \\
& ACA & 2 & 2 & 2 & 2 & 2 & 6 & 2 & 6 \\
\hline \hline
\end{tabular}
\centering \caption{Values of the coefficient multiplying $c_H$ ($K_H = 8\vec{ J_H^c} \cdot  \vec{J_H^a})$ and  $c_V $ $(K_V = 8\vec{ J_V^c} \cdot  \vec{J_V^a})$ computed  from the value of the tensor product of  the codon  $XYZ$ with the anticodon $NY_cX_c$,  for the quadruplets.} 
 \label{coef-q}
 \end{table}
 
 \clearpage
 
 \begin{table}[htbp]
\begin{tabular}{|c|c||c|c||c|c||c|c||c|c||} \hline 
a.a & codon &$ K_{H,C}$ & $ K_{V,C}$   &$ K_{H,U}$ & $ K_{V,U}$ &$ K_{H,G}$ & $ K_{V,G}$&   $ K_{H,A}$ & $ K_{V,A}$  \\
\hline \hline
His & CAC & 2 & 2 & -10 & 2 & 2 & -6 & -10 & -6 \\
& CAU & 2 & 2 & 6 & 2 & 2 & -6& 6 & -6 \\
\hline
Gln & CAG & 2 & 2 & -10 & 2 & 2 & 2 & 6 & 2 \\
& CAA & 2 & 2 & 6 & 2 & 2 & 2 & 6 & 2 \\
\hline
Phe & UUC & -10 & -10& -6 & -10& -10 & -30& -6 & -30 \\
& UUU & 6 & -10& 18 & -10& 6 & -30& 18 & -30 \\
\hline
Leu & UUG & -10 & -6& -6 & -6&-10 & -10& -6 & -10 \\
& UUA & 6 & -6 & 18 & -6 & 6 & -10 & 18 & -10 \\
\hline
Cys & UGC & 6 & 2 & -10 & 2 & 6 & -6 & -10 & -6 \\
& UGU & 2 & 2 & 2 & 2 & 2 & -6 & 2 & -6 \\
\hline
Trp & UGG & 6 & 2 & -10 & 2 & 6 & 2 & -10 & 2 \\
& UGA & 2 & 2 & 2 & 2 & 2 & 2 & 2 & 2 \\
\hline\hline
Tyr  & UAC & -10 & 2 & -6 & 2 & -10 & -6 & -6 & -6 \\
& UAU & 6 & 2 & 18 & 2 & 6 & -6 & 18 & -6 \\
\hline\hline
Asp & GAC & 2 & 18 & -10 & 18& 2 & -6& -10 & -6 \\
& GAU & 2 & 18& 6 & 18& 2 & -6& 6 & -6 \\
\hline
Glu & GAG & 2 & 18& -10 & 18& 2 & 18& -10 & 18 \\
& GAA & 2 & 18 & 6 & 18 & 2 & 18& 6 & 18 \\
\hline\hline
 Ile & AUC & -10 & 6& -6 & 6 & -10 & -22& -6 & -22 \\
& AUU & 6 & 6 & 18 & 6 & 6 & -22& 18 & -22 \\
\hline
Met & AUG & -10 & 2 & -6 & 2 &-10 & 6 & -6 & 6 \\
& AUA & 6 & 2 & 18 & 2 & 6 & 6 & 18 & 6 \\
 \hline \hline
Ser & AGC & 6 & 18 & -10 & 18 & 6 & -6 & -10 & -6 \\
& AGU & 2 & 18 & 2 & 18 & 2 & -6 & 2 & -6 \\
\hline\hline
Asn & AAC & -10 & 18 & -6 & 18& -10 & -6& -6 & -6 \\
& AAU & 6 & 18& 18 & 18 & 6 & -6& 18 & -6 \\
\hline
Lys & AAG & -10 & 18& -6 & 18& -10 & 18& -6 & 18 \\
& AAA & 6 & 18 & 18 & 18 & 6 & 18& 18 & 18 \\
 \hline \hline
\end{tabular}
\centering \caption{Values of the coefficient multiplying $c_H$ ($K_H = 8\vec{ J_H^c} \cdot  \vec{J_H^a})$ and  $c_V $ $(K_V = 8\vec{ J_V^c} \cdot  \vec{J_V^a})$ computed  from the value of the tensor product of  the codon  $XYZ$ with the anticodon $NY_cX_c$,  for the doublets.} 
 \label{coef-d}
 \end{table}
 
\clearpage

 \begin{table}[htbp]
\begin{tabular}{|c|c||c||c|} \hline 
a.a & anticodon & coeff. $ c_{H}$ &  coeff. $ c_{V}$  \\
\hline \hline
Pro & CGG &  18$P^q_S$ +6(1-$P^q_S$) & -10$P^q_Y$ - 6(1-$P^q_Y$)  \\
& UGG &  -6$P^q_S$ -10(1-$P^q_S$)& -10$P^q_Y$ - 6(1-$P^q_Y$) \\
& GGG &  18$P^q_S$ +6(1-$P^q_S$) & -30$P^q_Y$ - 10(1-$P^q_Y$) \\
& AGG &  -6$P^q_S$ -10(1-$P^q_S$& -30$P^q_Y$ - 10(1-$P^q_Y$)  \\
\hline 
Leu & CAG & 2 &   -10$P^q_Y$ - 6(1-$P^q_Y$)  \\
& UAG &   -10$P^q_S$ + 6(1-$P^q_S$) &  -10$P^q_Y$ - 6(1-$P^q_Y$)  \\
& GAG & 2 &  -30$P^q_Y$ - 10(1-$P^q_Y$) \\
& AAG &   -10$P^q_S$ + 6(1-$P^q_S$) &  -30$P^q_Y$ - 10(1-$P^q_Y$) \\
\hline
Arg & CCG & 18$P^q_S$ +6(1-$P^q_S$)  & 2   \\
& UCG & -6$P^q_S$ -10(1-$P^q_S$) & 2   \\
& GCG & 18$P^q_S$ +6(1-$P^q_S$) & -6$P^q_Y$ +2(1-$P^q_Y$)  \\
& ACG & -6$P^q_S$ - 10(1-$P^q_S$)&   -6$P^q_Y$ +2(1-$P^q_Y$)  \\
\hline
Ala  & CGC & 18$P^q_S$ + 6(1-$P^q_S$) &  6$P^q_Y$ + 2(1-$P^q_Y$)  \\
& UGC &  -6$P^q_S$ - 10(1-$P^q_S$) &  6$P^q_Y$ + 2(1-$P^q_Y$)   \\
& GGC &  18$P^q_S$ + 6(1-$P^q_S$) &  -22$P^q_Y$ + 6(1-$P^q_Y$)    \\
& AGC & -6$P^q_S$ - 10(1-$P^q_S$)  &   -22$P^q_Y$ + 6(1-$P^q_Y$)   \\
\hline
Gly  & CCC & 18$P^q_S$ + 6(1-$P^q_S$)  & 18  \\
& UCC &  -6$P^q_S$  -10(1-$P^q_S$)& 18  \\
& GCC & 18$P^q_S$ + 6(1-$P^q_S$) &   6$P^q_Y$  + 18(1-$P^q_Y$)  \\
& ACC &   -6$P^q_S$  -10(1-$P^q_S$) &  6$P^q_Y$  + 18(1-$P^q_Y$)  \\
\hline 
Val & CAC & 2 & 6$P^q_Y$  + 2(1-$P^q_Y$)  \\
&UAC & -10$P^q_S$ + 6(1-$P^q_S$) & 6$P^q_Y$  + 18(1-$P^q_Y$)   \\
& GAC & 2 &  -22$P^q_Y$ + 6(1-$P^q_Y$)    \\
& AAC & -10$P^q_S$ + 6(1-$P^q_S$) &  -22$P^q_Y$ + 6(1-$P^q_Y$)   \\
\hline
Ser & CGA & 6$P^q_S$ + 2(1-$P^q_S$) &  -10$P^q_Y$ - 6(1-$P^q_Y$) \\
& UGA &  -10$P^q_S$ + 2(1-$P^q_S$)  &   -10$P^q_Y$ - 6(1-$P^q_Y$) \\
& GGA &  6$P^q_S$ + 2(1-$P^q_S$)  &  -30$P^q_Y$ - 10(1-$P^q_Y$)\\
& AGA & -10$P^q_S$ + 2(1-$P^q_S$)  & -30$P^q_Y$ - 10(1-$P^q_Y$)    \\
\hline
Thr  & CGU &  6$P^q_S$ +2(1-$P^q_S$) & 6$P^q_Y$ + 2(1-$P^q_Y$)  \\
& UGU &  -10$P^q_S$ +2(1-$P^q_S$) & 6$P^q_Y$ + 2(1-$P^q_Y$)  \\
& GGU &   6$P^q_S$ +2(1-$P^q_S$)  & -22$P^q_Y$ + 6(1-$P^q_Y$)  \\
& AGU & -10$P^q_S$ +2(1-$P^q_S$)  & -22$P^q_Y$ + 6(1-$P^q_Y$)  \\
\hline \hline
\end{tabular}
\centering \caption{Value of the coefficients multiplying   $c_H $ and  $c_V $ in ${\cal T}_{av}$ , computed for any anticodon and averaged over the four codons for each quadruplet.} 
 \label{avT-q}
 \end{table}

\clearpage

\begin{longtable}{|c|c||c||c|} 
 
 \hline 
\multicolumn{1}{|c|} {\textbf a.a} &  \multicolumn{1} {|c|} {\textbf anticodon} & \multicolumn{1} {|c|}  {\textbf coeff. $ c_{H}$} &  
\multicolumn{1}{|c|} {\textbf  coeff. $ c_{V}$} \\
\hline \hline
\endfirsthead
\hline
\multicolumn {3}{l}  {\textbf Continued from previous page} \\ 
\hline
\multicolumn{1}{|c|} {\textbf a.a} &  \multicolumn{1} {|c|} {\textbf anticodon} & \multicolumn{1} {|c|}  {\textbf coeff. $ c_{H}$} &  
\multicolumn{1}{|c|} {\textbf  coeff. $ c_{V}$} \\
 \endhead
 
\multicolumn{2}{r}{\textbf Continued on next page} \\ \hline

\endfoot
\endlastfoot
His & CUG &  2 & 2  \\
& UUG &  -10$P^d_C$ + 6(1-$P^d_C$)& 2\\
& GUG & 2 & -6 \\
& AUG & -10$P^d_C$ + 6(1-$P^d_C$)& -6 \\
\hline  \hline
Gln & CUG &  2 & 2  \\
& UUG &  -10$P^d_G$ + 6(1-$P^d_G$)& 2\\
& GUG & 2 & 2 \\
& AUG &  6 & 2 \\
\hline\hline
Phe & CAA &  -10$P^d_C$ + 6(1-$P^d_C$) & -10  \\
& UAA &  -6$P^d_C$ + 18(1-$P^d_C$)& -10\\
& GAA & -10$P^d_C$ + 6(1-$P^d_C$) & -30 \\
& AAA & -6$P^d_C$ + 18(1-$P^d_C$)& -30 \\
\hline  \hline
Leu & CAA &  -10$P^d_G$ + 6(1-$P^d_G$) & -6  \\
& UAA &  -6$P^d_G$ + 18(1-$P^d_G$)& -6\\
& GAA & -10$P^d_G$ + 6(1-$P^d_G$) & -10 \\
& AAA & -6$P^d_G$ + 18(1-$P^d_G$)& -10 \\
\hline \hline
Cys & CCA &  6$P^d_C$ + 2(1-$P^d_C$) & 2  \\
& UCA &  -10$P^d_C$ + 2(1-$P^d_C$)& 2\\
& GCA & 6$P^d_C$ + 2(1-$P^d_C$) & -6 \\
& ACA & -10$P^d_C$ + 2(1-$P^d_C$)& -6 \\
\hline  \hline
Trp & CCA &  6$P^d_G$ + 2(1-$P^d_G$) & 2  \\
& UCA &  -10$P^d_G$ + 2(1-$P^d_G$)& 2\\
& GCA & 6$P^d_G$ + 2(1-$P^d_G$) & 2 \\
& ACA & -10$P^d_G$ + 2(1-$P^d_G$)& 2 \\
\hline \hline
Tyr & CUA &  -10$P^d_C$ + 6(1-$P^d_C$) & 2  \\
& UUA &  -6$P^d_C$ + 18(1-$P^d_C$)& 2\\
& GUA & -10$P^d_C$ + 6(1-$P^d_C$) & -6 \\
& AUA & -6$P^d_C$ + 18(1-$P^d_C$)& -6 \\
\hline  \hline
Ser & CCU &  6$P^d_C$ + 2(1-$P^d_C$) & 18  \\
& UCU &  -10$P^d_C$ + 2(1-$P^d_C$)& 18\\
& GCU & 6$P^d_C$ + 2(1-$P^d_C$) & -6 \\
& ACU & -10$P^d_C$ + 2(1-$P^d_C$)& -6 \\
\hline  \hline
Asp & CUC & 2 & 18  \\
& UUC &  -10$P^d_C$ + 6(1-$P^d_C$)& 18\\
& GUC &  2 & -6 \\
& AUC & -10$P^d_C$ + 6(1-$P^d_C$)& -6 \\
\hline  \hline
Glu & CUC &   2 & 18  \\
& UUC &  -10$P^d_G$ + 6(1-$P^d_G$)& 18\\
& GCA & 2 & 18 \\
& ACA & -10$P^d_G$ + 6(1-$P^d_G$)& 18 \\
\hline \hline
Ile & CAU &  -10$P^d_C$ + 6(1-$P^d_C$)& 6  \\
& UAU &  -6$P^d_C$ + 18(1-$P^d_C$)& 6\\
& GAU & -10$P^d_C$ + 6(1-$P^d_C$ & -22 \\
& AAU & -6$P^d_C$ + 18(1-$P^d_C$)& -22 \\
\hline  \hline
Met & CAU &   -10$P^d_G$ + 6(1-$P^d_G$)& 2  \\
& UAU &  -6$P^d_G$ +186(1-$P^d_G$)& 2\\
& GAU &  -10$P^d_G$ + 6(1-$P^d_G$)& 6 \\
& AAU &   18$P^d_G$ + 6(1-$P^d_G$) & 6 \\
\hline\hline
Asn & CUU &  -10$P^d_C$ + 6(1-$P^d_C$) & 18  \\
& UUU &  -6$P^d_C$ + 18(1-$P^d_C$)& 18\\
& GUU & -10$P^d_C$ + 6(1-$P^d_C$) & -6 \\
& AUU & -6$P^d_C$ + 18(1-$P^d_C$)& -6 \\
\hline  \hline
Lys & CUU &  -10$P^d_G$ + 6(1-$P^d_G$) & 18  \\
& UUU &  -6$P^d_G$ + 18(1-$P^d_G$)& 18\\
& GUU & -10$P^d_G$ + 6(1-$P^d_G$) & 18 \\ 
& AUU & -6$P^d_G$ + 18(1-$P^d_G$)& 18 \\
\hline \hline
\caption{ Value of the coefficients multiplying   $c_H $ and  $c_V $ in ${\cal T}_{av}$ , computed for any anticodon and averaged over the two codons for each doublet.} 
\label{avT-d2}
 \end{longtable}

\newpage

\begin{table}[htbp]
\begin{tabular}{|c||c||c||c|} \hline 
a.a & sign $ c_{H}$ & anticodon & note   \\
\hline \hline
His &  - & GUG & $P^d_C > 0,25$ \\
\hline  
Gln & + &  UUG &  $P^d_G > 0,25$   \\
\hline\hline
Phe & - &   GAA  &  \\
\hline  
Leu & + &   UAA &  \\
\hline \hline
Cys &+ &   GCA &   \\
 \hline  
Trp & - & UCA &   \\
\hline \hline
Tyr &  - & GUA & \\
\hline  \hline
Ser & + &  GCU &  \\
\hline  \hline 
Asp & + & GUC &  $P^d_C > 0,25$ \\
\hline  
Glu &- & UUC &    $P^d_G > 0,25$ \\
\hline \hline
Ile &  + & GAU  & \\
\hline  
Met  & - &  CAU &   \\
\hline\hline
Asn & - &   GUU &  \\
\hline  
Lys & + &   UUU &  \\
\hline \hline
\end{tabular}
\centering \caption{Anticodon minimizing  the operator ${\cal T}$, averaged over the two codons,  for any amino acid encoded by a doublet, specifying the sign of $c_H$. } 
 \label{min-d}
 \end{table}

\newpage

\begin{table}[htbp]
\begin{tabular}{|c||c||c||} \hline 
a.a & codon &  anticodon    \\
\hline \hline
Pro & CCG & CGG    \\
\hline \hline  
Leu & {\bf CUA} &   {\bf UAG}  \\
\hline\hline
Arg & CGG & CCG    \\
\hline \hline 
Ala & GCG & CGC \\
\hline \hline
Val & {\bf GUA} & {\bf UAC}  \\
 \hline \hline  
Gly & GGG  & CCC    \\
\hline \hline
Ser & UCG &  CGA \\
\hline  \hline
Thr & ACG  & CGU  \\
\hline  \hline
His &  CAC & GUG \\
\hline  
Gln & CAG & CUG \\
\hline\hline
Phe   & UUC  & GAA \\
\hline \hline
Cys  & {\bf UGU} &  {\bf ACA}   \\
 \hline  
Trp & {\bf UGA} &  {\bf  UCA}  \\
\hline \hline
Tyr & UAC  &  GUA \\
\hline  \hline 
Asp & GAC &  GUC \\
\hline  
Glu  & GAG & CUC  \\
\hline \hline
Ile  & AUC & GAU \\
\hline  
Met  & AUG &  CAU  \\
\hline\hline
Asn & AAC & GUU \\
\hline  
Lys & AAG & CUU \\
\hline \hline
\end{tabular}
\centering \caption{The couple of codon-anticodon which minimizes the operator ${\cal T}$
 with $c_V < 0$ and $c_H < 0$ for the strong dinucleotides (first 8 rows) and  $c_V$ underdetermined  $c_H > 0$  for the weak dinucleotides 
 in the Ancient Genetic Code.}
 \label{codacod-agc}
 \end{table}

\begin{table}[htbp]
\begin{tabular}{|c||c||c||c||c||c|} \hline 
a.a & codon & anticodon &  sign $ c_{H}$ &  sign $ c_{V}$ & note   \\
\hline \hline
Pro & CCN & UGG  & + &  - & \\
\hline \hline  
Leu & CUN & UAG &  + &  -  & \\
\hline\hline
Arg & CGN & UCG & + &  - &  \\
\hline \hline 
Ala & GCN & UGC  &+ & - &  \\
\hline \hline
Val & GUN & UAC  & +  &  -  &   \\
 \hline \hline  
Gly & GGN  & UCC & + & - &   \\
\hline \hline
Ser & UCN & UGA &  + & - & \\
\hline  \hline
Thr & ACN  & UGU  & + &  - &  \\
\hline  \hline
His/Gln & CAN & UUG  & + &  - & $P_S  >  1/4 $\\
\hline\hline
Phe/Leu' & UUN & UAA &  - &  - &  \\
\hline \hline
Cys/Trp & UGN & UCA & + &  + &   \\
\hline \hline
Tyr & UAY & GUA &  + & + & \\
\hline  \hline 
Asp/Glu & GAN & UUC &  + & - & $P_S  >  1/4 $  \\
\hline \hline
Ile/Met & AUN & UAU &  - & -  & $P_Y  >  1/8 $\\
\hline\hline
Asn/Lys & AAN & UUU & - &   - &  \\
\hline\hline
Ser'/Arg' & AGN & UCU &  + &   - &  \\
\hline \hline
\end{tabular}
\centering \caption{Sign of coupling constants  minimizing the operator ${\cal T}$, averaged over the codons,  for any amino acid encoded in the Archetypal Genetic Code. We denote by a prime 
the a.a. encoded by  the sub-part of the sextet corresponding to a doublet.} 
 \label{sign-arco}
 \end{table}

\begin{table}[htbp]
\begin{tabular}{|c||c||c||c||c||c|} \hline 
a.a & codon & anticodon &sign $ c_{H}$ &  sign $ c_{V}$ & note   \\
\hline \hline
Pro & CCN & UGG &  + &  - & \\
\hline \hline  
Leu & CUN & UAG & + &  -  & $P_S > 1/4$ \\
\hline\hline
Arg & CGN & UCG & + &  - &  \\
\hline \hline 
Ala & GCN & UGC &  + & - &  \\
\hline \hline
Val & GUN & UAC & +  &  -  & $P_S > 1/4$   \\
 \hline \hline  
Gly & GGN  & UCC & + & - &   \\
\hline \hline
Ser & UCN & UGA &  + & - & \\
\hline  \hline
Thr & ACN  & UGU &  + &  - &$P_Y >  1/8$ \\
\hline  \hline
His & CAY & GUG &  + &  + & $P_C >  3/8$ \\
\hline
Gln & CAR & UUG & - &  und. & $P_G  <  1/4$ \\
\hline\hline
Phe	 & UUY & GAA & + &  + &  \\
\hline
Leu'	 & UUR & UAA &  - &  und. &  \\
\hline \hline
Cys  & UGY & GCA & - &  + &   \\
\hline
Trp & UGR & {\bf UCA} & + &  und. &   \\
\hline \hline
Tyr & UAY & GUA & + & + & \\
\hline  \hline 
Asp  & GAY & GUC & + & + & $P_C > 1/4$  \\
\hline
Glu & GAR & UUC & - & und. & $P_G < 1/4$   \\
\hline \hline
Ile & AUY &  GAU &  + & +  & \\
\hline
Met & AUR &  UAU & - & und.  & \\
\hline\hline
Asn & AAY & GUU &  + &   + &  \\
\hline
Lys & AAR & UUU & - &   und. &  \\
\hline \hline
Ser' & AGY & GCU &  - &   + &  \\
\hline
Arg' & AGR & UCU & + &   und. &  \\
\hline \hline
\end{tabular}
\centering \caption{Sign of coupling constants  minimizing the operator ${\cal T}$, averaged over the codons,  for any amino acid encoded in the Early Genetic Code.}
 \label{sign-eco}
 \end{table}

\begin{table}[htbp]
\footnotesize
\begin{tabular}{|c||c|c||c|c||c|c|} \hline 
a.a. & $ c_{H}$ Ancient &   $ c_{V}$ Ancient  &  $ c_{H}$ Archet. &   $ c_{V}$ Archet &  $ c_{H}$ Early &   $ c_{V}$ Early    \\
\hline \hline
 Pro &  - &  - & + &  - & + &  -   \\
\hline \hline  
 Leu & -  &  -  & + &  - & + &  -  \\
\hline\hline
 Arg & - &  - & + &  - & + &  -   \\
\hline \hline 
 Ala & - & - & + &  - & + &  -  \\
\hline \hline
 Val & -  &  -  &  + &  - & + & - \\
 \hline \hline  
 Gly  &  -& - & + &  - & + &  -     \\
\hline \hline
 Ser  & -  & - & + &  - & + &  -  \\
\hline  \hline
 Thr  & - &  - & + &  - & + &  -    \\
\hline  \hline\hline
His & + & UND.  & + &  + & + &  +   \\
\hline  
Gln & + &  UND.    & + &  - & - &   UND.     \\
\hline  \hline
Phe & + & UND.    & - &  - & + &  +    \\
\hline 
Leu' &  ===  &  ===  & - &  - & - &  UND.    \\
\hline  \hline
Cys & + & UND.    & + &  + & - &  +    \\
\hline 
Trp & + &  UND.    & + &  + & + &  UND.    \\
\hline  \hline
Tyr & + & UND.     & + &  + & + &  +   \\
\hline  \hline
Asp & + &   UND.  & + &  - & + &  +   \\
\hline   
Glu & + &  UND.   & + &  - &  - &  UND.     \\
\hline  \hline
Ile & + & UND.    & - &  - & + &  +  \\
\hline   
Met & + & UND.    & - &  - & - &  UND.    \\
\hline  \hline
 Asn & + &  UND.   & - &  - & + &  +   \\
\hline 
Lys & + &    UND.      & - &  - & - &  UND.   \\
\hline  \hline
Ser' &   ===  &  ===   & + &  - & - & +   \\
\hline   
Arg' &$\Relbar\Relbar\Relbar$    &  ===   & + &  - & + & UND.   \\
\hline  \hline
\end{tabular}
\centering \caption{Sign of coupling constants  in the different genetic code. We denote with a  prime (`` '  '')  the amino-acid encoded by the doublet sub-part of a sextet. The symbol $\Relbar\Relbar\Relbar$  and UND. indicate,  respectively, that the concerned amimo-acid is not present in the considered genetic code and that the sign is undetermined. } 
 \label{sign-evol}
 \end{table}

\newpage

\begin{figure}[htbp]
\includegraphics[width=16cm]{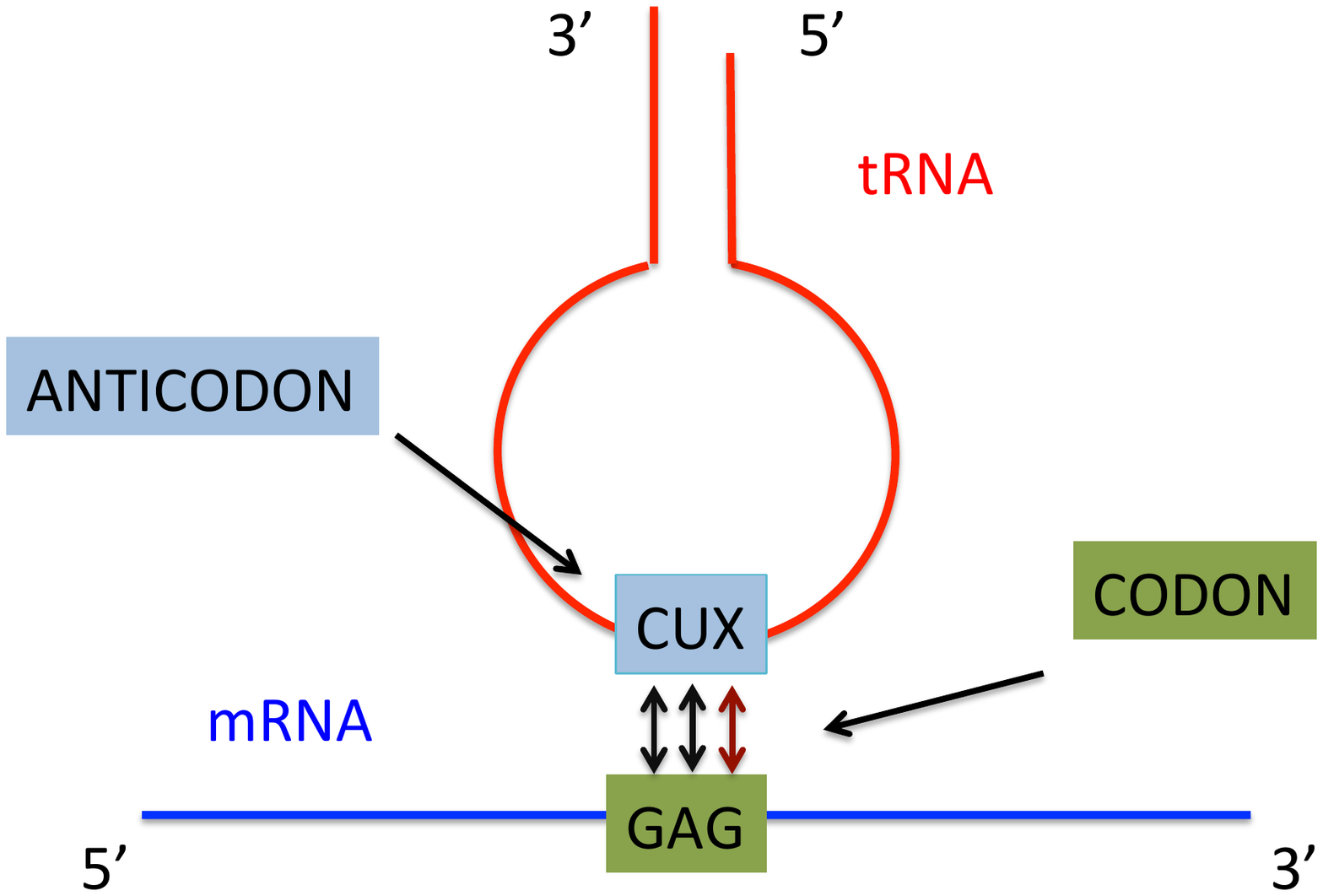}
\centering\caption{Anticodon-codon reading scheme.}
 \label{Reading.pdf}
 \end{figure}

\newpage

\begin{figure}[htbp]
\includegraphics[width=16cm]{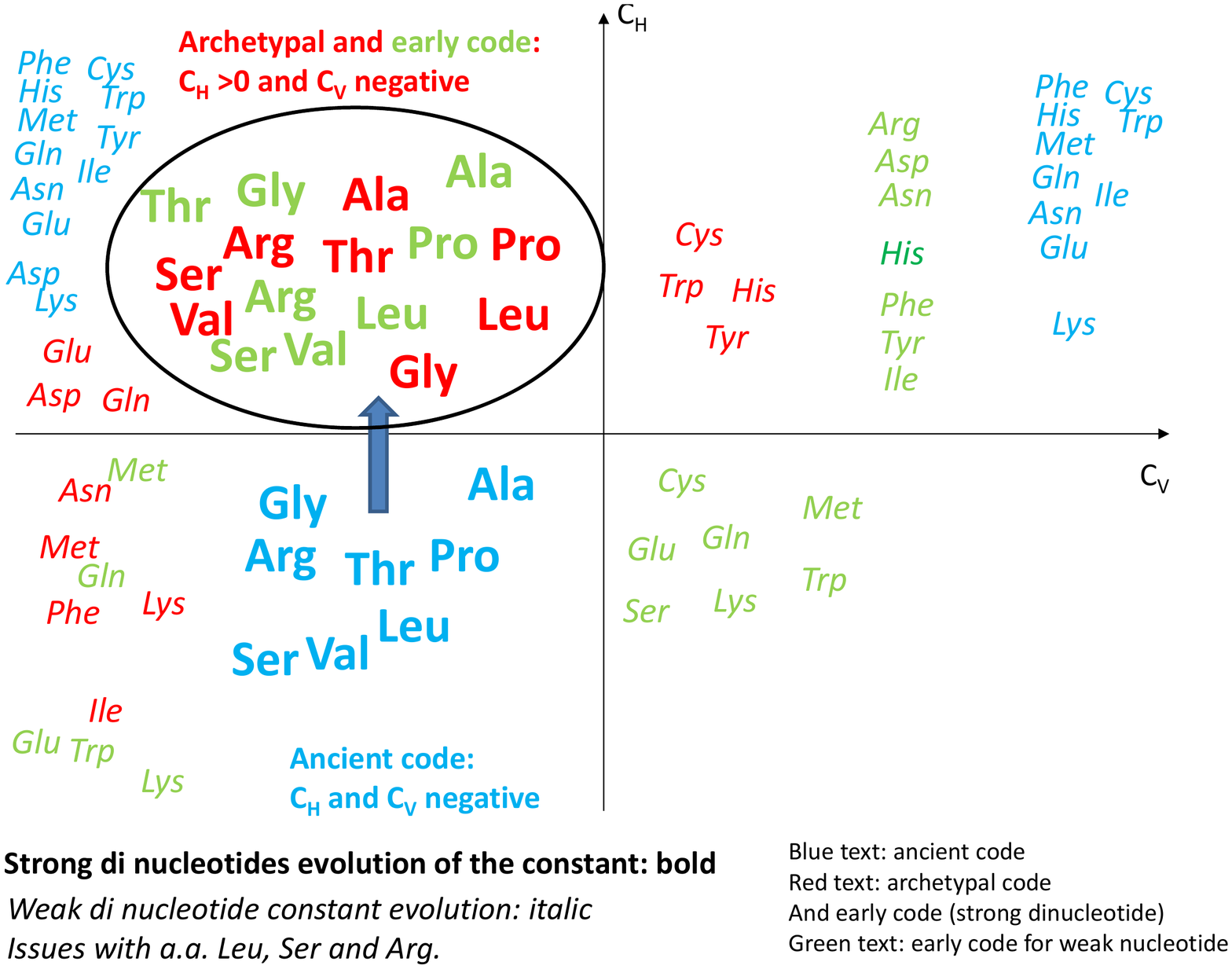}
\centering\caption{Sign of the constant $c_H$ and $c_V$ in the evolution of the genetic code.}
 \label{cH.pdf}
 \end{figure}

\end{document}